\documentstyle[12pt]{article}
\title{Relativistic contraction and related effects in noninertial frames}
\author{Hrvoje Nikoli\'c  \\
Theoretical Physics Division, Rudjer Bo\v{s}kovi\'{c} Institute, \\
P.O.B. 1016, HR-10001 Zagreb, Croatia \\
{\normalsize hrvoje@faust.irb.hr} \\
\makebox[1in]{} \\
}
\date{\today}
\begin{document}
\maketitle
\begin{abstract}
 Although there is no relative motion among different points on a 
 rotating disc, each point belongs to a {\em different} noninertial 
 frame. This fact, not recognized in previous approaches to 
 the Ehrenfest paradox and related problems, is exploited to give 
 a correct treatment of a rotating ring and a rotating disc.  
 Tensile stresses are recovered, but, contrary to the 
 prediction of the standard approach, it is found that an 
 observer on the rim of the disc will see equal lengths of other differently 
moving objects   
 as an inertial observer whose instantaneous 
 position and velocity are equal to that of the observer on the rim. 
The rate of clocks at various positions, as seen by various 
observers, is also discussed. 
 Some results are generalized for observers arbitrarily moving in  
 flat or curved spacetime. 
 The generally accepted formula for the space line element 
 in a non-time-orthogonal frame is found 
 inappropriate in some cases.  
 Use of Fermi coordinates leads to the result that for any observer 
the velocity of light is isotropic and is
equal to $c$, providing that it is measured by propagating a light   
beam in a small neighborhood of the observer.
\end{abstract}  
\vspace*{0.5cm}
PACS number(s): 03.30.+p, 04.20.Cv

\section{Introduction}

If a body moves with a constant velocity, then, as is well known, 
the body is relativistically contracted in the direction of motion, 
whereas its length in the normal direction is unchanged. A naive  
generalization to a rotating disc leads to the conclusion that the 
circumference of the disc is contracted, whereas the radius of the 
disc is unchanged. This paradox is known as the Ehrenfest paradox. 
Obviously, the paradox is a consequence of the application of the 
constant-velocity result to a system with a nonconstant velocity.
  
The standard resolution (see \cite{gron1}, \cite{gron2} and references 
therein)  
of the Ehrenfest paradox is as follows: One introduces the coordinates 
of the rotating frame $S'$ 
\begin{equation}\label{eq1}
 \varphi'=\varphi -\omega t \; , \;\;\;\; r'=r \; , \;\;\;\; z'=z 
  \; , \;\;\;\; t'=t \; ,
\end{equation}
where $\varphi$, $r$, $z$, $t$ are cylindrical coordinates of the 
inertial frame $S$ and $\omega$ is the angular velocity. The metric in $S'$ is 
given by
\begin{equation}\label{eq2}
 ds^2=(c^2-\omega^2 r'^2)dt'^2 -2\omega r'^2 \, d\varphi' dt' -dr'^2 
 -r'^2 \, d\varphi'^2 -dz'^2 \; .
\end{equation}
It is generally accepted that the space line element should be 
calculated by the formula \cite{land}
\begin{equation}\label{eq3}
 dl'^2=\gamma'_{ij}dx'^i dx'^j \; , \;\;\; i,j=1,2,3 \; ,
\end{equation}
where
\begin{equation}\label{eq4}
 \gamma'_{ij}=\frac{g'_{0i}g'_{0j}}{g'_{00}}-g'_{ij} \; .
\end{equation}
This leads to the circumference of the disc 
\begin{equation}\label{eq5}
 L'=\int_{0}^{2\pi} \frac{r' d\varphi'}{\sqrt{1-\omega^2 r'^2/c^2}} 
   =\frac{2\pi r'}{\sqrt{1-\omega^2 r'^2/c^2}}   
   \equiv\gamma(r')2\pi r' \; . 
\end{equation} 
The circumference of the same disc as seen from $S$ is $L=2\pi r=2\pi r'$. 
If the disc is constrained  to have the same radius $r$ as the same 
disc when it does not rotate, then $L$ is not changed by the rotation, but 
the proper circumference $L'$ is larger than the proper circumference of 
the nonrotating disc. This implies that there are tensile stresses in the 
rotating disc.

However, there is something wrong with this standard resolution 
of the Ehrenfest paradox. Consider a slightly simpler situation; 
a rotating ring in a rigid nonrotating 
circular gutter with the radius $r=r'$.  
The statement that (\ref{eq5}) represents the 
proper circumference implies that the {\em proper} frame of the 
rotating ring is given by (\ref{eq1}). This
means that an observer on the ring sees that the 
circumference is $L'=\gamma L$. The circumference of the gutter 
seen by him 
cannot be different from the circumference of the ring
seen by him, so the 
observer on the ring sees that the circumference of the 
relatively moving gutter is {\em larger} than the proper 
circumference of the gutter, whereas we expect that he should 
see that it is smaller. 
This leads to another paradox. It cannot be resolved by saying that 
the observer on the ring accelerates, because one can consider a limit 
$r\rightarrow\infty$, $\omega\rightarrow 0$, $r\omega\equiv u=$constant, 
which implies that the acceleration $a=r\omega^2$ becomes zero, 
whereas the paradox remains.     

Before explaining how we resolve this paradox, we give some 
general notes on the physical meaning of various coordinate frames in 
the theory of relativity. In practice, one usually uses the  
coordinates that simplify the technicalities of the 
physical problem considered. 
For example, when one describes physical effects in a 
rigid body, it may be convenient to use a comoving coordinate frame, 
i.e., a frame in which all particles of the rigid body have 
constant spatial coordinates. The coordinates of $S'$ in 
(\ref{eq1}) may be interpreted in this way.  
However, {\em the choice of the  
coordinate frame is more than a matter of convenience}.  
The main lesson we have learned from Lorentz 
coordinate frames is the fact that what an observer observes  
(time intervals, space intervals, components of a tensor, ...) 
depends on how the observer moves. The main purpose of 
theoretical physics is to predict what will be {\em observed} under 
given circumstances. Therefore, unless stated otherwise, 
in this paper {\em by a 
coordinate frame we understand a coordinate frame that 
is inherent to an observer}, not to a set of physical particles.  
Our criticism of some earlier treatments originates from  
such an interpretation of coordinate frames. To avoid a possible 
misunderstanding, we note that coordinate frames do not 
necessarily need to be interpreted in this way, in which case  
our criticism does not apply.  

We resolve the paradox by recognizing that, 
according to our interpretation, the frame defined by 
(\ref{eq1}) is the proper frame {\em only} of the observer at 
$r=r'=0$. This observer has no velocity relative to $S$, 
so the corresponding coordinate transformation (\ref{eq1}) 
does not depend on any velocity. As will become clear from the discussion  
of Section 2, the frame defined by (\ref{eq1}) is actually the Fermi 
frame of an observer who rotates, but has no velocity with respect 
to the frame $S$.  
Observers at different positions 
on the rotating disc have different velocities, so one has to use a 
different coordinate transformation for each of them. In other 
words, {\em although there is no relative motion among different points on a
rotating disc, each point belongs to a different noninertial
frame.} 
This is not strange to those who 
are familiar with the theory of Fermi coordinates \cite{mtw}, \cite{synge}, 
but it seems that many relativity-theorists are not. 
%To rephrase this crucial fact, there is no such thing as a
%``frame of a rotating disc as a whole", simply because each part 
%of the disc belongs to a different frame. A disc is nothing else but 
%a set of many interacting particles, each having its own trajectory. 
%Only if the axes of $S'$ in (\ref{eq1}) were realized as {\em material}
%curves, by, for example, drawing on the disc,    
%it would be meaningful to interpret them  
%as something inherent to the disc as a whole. 
%However, in the theory of relativity, the coordinate axes are not 
%material objects, but a purely mathematical construction. In particular, 
%it implies that $r'$ can be arbitrarily large in (\ref{eq2}), although 
%there is an coordinate singularity at $r'=c/\omega$. 

Note also that since we do not interpret the coordinates of $S'$ in
(\ref{eq1}) as something inherent to the disc as a whole, $r'$ can be
arbitrarily large in (\ref{eq2}), although
there is a coordinate singularity at $r'=c/\omega$. 
It resembles 
the Schwarzschild singularity of a black hole, where the radial coordinate 
is not restricted to be 
larger than the Schwarzschild radius. However, to avoid a possible 
misunderstanding, note that the coordinate singularity in (\ref{eq2}) does 
not correspond to an event horizon, because a rotating observer at 
$r'=0$ {\em can} receive information from $r'\geq c/\omega$.   

There is also another paradox connected with the standard approach to 
rotating frames. Let us consider how the nonrotating gutter looks like to  
a rotating observer in the center. His proper frame 
{\em is} given by (\ref{eq1}). If (\ref{eq3}) is the correct definition 
of the space line element, then he should see that the circumference 
of the gutter is larger 
than the proper circumference of the gutter by a factor $\gamma(r')$. 
However, $\omega r'/c$ can be 
arbitrarily large, so $\gamma(r')$ can be not only 
arbitrarily large, but also even imaginary. On the other hand, we know from 
everyday experience that the apparent velocity $\omega r'$ of stars,
due to our rotation,  
can exceed the velocity of light, but we see neither a contraction, 
nor an elongation of the stars observed. 

We resolve this paradox by examining the assumptions 
under which formula (\ref{eq3}) is obtained. We find that this formula 
should be used with great care and show that it is not applicable
in our case.
The correct definition of the space line element depends 
on how it is measured, and we find that, in our case, $\gamma'_{ij}$ should be 
replaced by $-g'_{ij}$ in (\ref{eq3}).  
 
It is fair to note that there are also some other ``nonstandard" 
approaches to the Ehrenfest paradox (see \cite{tart2}, \cite{kla},  
and references 
therein), but none of these approaches is similar to ours. 
In particular, the crucial fact that 
each point of the rotating ring belongs to a different frame has 
not been taken into account in any of these approaches. 
Formula (\ref{eq3}) has already been criticized \cite{kla}, but our criticism 
of (\ref{eq3}) is quite different and more general.  

The paper is organized as follows: In Section 2 we find the correct 
coordinate transformation that leads to the 
frame of an observer moving in flat spacetime. 
In Section 3 we explain why (\ref{eq3}) 
is not always a correct definition of a space line element and 
show that in a frame that corresponds to an observer in flat spacetime 
it is more appropriate to calculate the space line element by 
$-g'_{ij}$. We also make some general remarks on the physical meaning of 
general coordinate transformations.    
In Section 4 we study the relativistic contraction as seen by various 
observers and resolve the Ehrenfest paradox. In Section 5 we study the rate 
of clocks at various positions, as seen by various observers. 
In Section 6 we discuss the velocity of light as seen by various observers.
In Section 7 we discuss our results, resolve some additional physical 
problems, and give some generalizations. Section 8 is 
devoted to concluding remarks, where the relevance of our results 
to general relativity is emphasized.    

\section{The frame of an observer moving in flat spacetime}

The generalized Lorentz transformations for a local Fermi 
frame of an observer 
that has arbitrary time-dependent velocity and angular velocity 
in flat spacetime are found in 
\cite{nels}. We present the final results, using 
slightly different notation. Let $S$ be an inertial frame  
and let $S'$ be the frame of the observer whose velocity and angular velocity 
are $u^i(t')$ and $\omega^i(t')$, respectively, as seen by an observer in $S$. 
The coordinate transformation between these two frames is given by 
\begin{equation}\label{er1}
 x^i =-A_{j}^{\; i}(t')x'^j +\int_{0}^{t'} \gamma(t')u^i(t')\, dt' +
 \frac{1}{\mbox{\bf{u}}^2(t')} [\gamma(t')-1][u^k(t')A_{jk}(t')x'^j]u^i(t') \; ,
\end{equation}
\begin{equation}\label{er2}
 t=\int_{0}^{t'} \gamma(t')\, dt' +
 \frac{1}{c^2}\gamma(t')[u^k(t')A_{jk}(t')x'^j] \; ,
\end{equation}
where $\gamma(t')=1/\sqrt{1-\mbox{\bf{u}}^2(t')/c^2}$ and
$A_{ji}(t')=-A_{j}^{\; i}(t')$ is the 
rotation matrix evaluated at $\mbox{\bf{x}}'=0$. The rotation matrix  
satisfies the differential equation
\begin{equation}\label{er4}
 \frac{d A_{ij}}{dt}=-A_{i}^{\; k}\omega_{kj} \; ,
\end{equation}    
where $\omega_{ik}=\varepsilon_{ikl}\omega^{l}$, $\varepsilon_{123}=1$.  
The metric tensor in $S'$ is
\begin{eqnarray}\label{metric}
 & g'_{ij}=-\delta_{ij} \; , \;\;\;\;\;
   g'_{0j}=-(\mbox{\boldmath $\omega$}'\times\mbox{\bf{x}}')_j \; , &
\nonumber \\
 & g'_{00}=c^2 \left(
1+\displaystyle\frac{\mbox{\bf{a}}'\cdot\mbox{\bf{x}}'}{c^2}
    \right)^2 -(\mbox{\boldmath $\omega$}'\times\mbox{\bf{x}}')^2 \; , &
\end{eqnarray}
where
\begin{equation}
 \omega'^i =\gamma (\omega^i -\Omega^i) \; , \;\;\;\;\;
 a'^i =\gamma^2 \left[a^i +\frac{1}{\mbox{\bf{u}}^2}(\gamma
  -1)(\mbox{\bf{u}}\cdot\mbox{\bf{a}})u^i\right] \; ,
\end{equation}
$\Omega^{i}$ is the time-dependent Thomas precession frequency
\begin{equation}
\Omega_{i}=\frac{1}{2\mbox{\bf{u}}^2}(\gamma -1)\varepsilon_{ikj}
 (u^k a^j -u^j a^k) \; , 
\end{equation}
and $a^i=du^i/dt$ is the time-dependent acceleration.
The transformations (\ref{er1})-(\ref{er2}) are chosen such that 
the space origins of $S$ and $S'$ coincide for $t=t'=0$. If $\mbox{\bf{u}}$ 
is time independent and $\mbox{\boldmath $\omega$}=0$, then
(\ref{er1})-(\ref{er2}) reduce to the well-known ordinary Lorentz 
boosts. If $\mbox{\bf{u}}=0$ and $\mbox{\boldmath $\omega$}$ 
is time independent, then (\ref{er1})-(\ref{er2}) reduce to (\ref{eq1}). 
It is important to emphasize that $\mbox{\bf{u}}(t')$ is the velocity of the 
{\em space origin} $\mbox{\bf{x}}'=0$ of $S'$. If $S'$ is a rotating frame, then 
other space points of $S'$ have a different velocity. (Remind that rotation 
is {\em not} a motion along a circle, but rather a change of orientation 
of the axes with respect to an inertial frame.) Therefore, 
in general, $S'$ is the proper frame {\em only}  
of the observer at $\mbox{\bf{x}}'=0$. 
Note also that $g'_{\mu\nu}=\eta_{\mu\nu}$ only at $\mbox{\bf{x}}'=0$, 
which is another confirmation that $S'$ is the frame of the observer at
$\mbox{\bf{x}}'=0$ only. The metric (\ref{metric}) is also consistent 
with a more general theory of Fermi coordinates \cite{mtw},  
which are coordinates of an observer arbitrarily moving in curved spacetime, 
and also have the property that $g_{\mu\nu}=\eta_{\mu\nu}$ at the 
space origin, i.e., at the position of the observer. Note also that if 
$\mbox{\bf{a}}'$ and $\mbox{\boldmath $\omega$}'$ vanish, then (\ref{metric}) 
is a metric of an inertial frame and is equal to $\eta_{\mu\nu}$ 
everywhere, so, in this case, $S'$ can be considered as a frame of an observer at 
{\em arbitrary} constant $\mbox{\bf{x}}'$. 

It is interesting to note that the geometrical
construction of Fermi coordinates is well established \cite{mtw},
\cite{synge}, but no analog of (\ref{er1})-(\ref{er2}) is
known for curved spacetime. The transformations (\ref{er1})-(\ref{er2})
are obtained by summation of infinitesimal Lorentz transformations (and
rotations). It is not so easy to find an analog of Lorentz transformations
in curved spacetime, because they correspond to the coordinate
transformation between Fermi frames of two different free-falling observers. 
We can, however,  
write the transformations (\ref{er1})-(\ref{er2}) 
in a more elegant form, which could be illuminating 
for a generalization to curved spacetime. Let 
\begin{equation}\label{el1}
x^{\mu}=f^{\mu}(t',\mbox{\bf{x}}';\mbox{\bf{u}}) 
\end{equation}
denote the ordinary Lorentz transformations, i.e., the transformations 
between two inertial frames specified by the relative velocity
$\mbox{\bf{u}}$, which can be considered as the relative velocity 
between two inertial (free-falling) 
observers at the instant when they have the same position. The differential 
of (\ref{el1}) is 
\begin{equation}\label{el2}
dx^{\mu}=f^{\mu}_{\; ,\nu} (t',\mbox{\bf{x}}';\mbox{\bf{u}}) dx^{\nu} \; .
\end{equation}
The transition to a noninertial frame introduces a time-dependent 
velocity: $\mbox{\bf{u}} \rightarrow \mbox{\bf{u}}(t')$. 
The transformations (\ref{er1})-(\ref{er2}) may be obtained by integrating 
(\ref{el2}) in the following way:
\begin{equation}\label{el3}
x^{\mu}=\int_{0}^{t'}f^{\mu}_{\; ,0} (t',0;\mbox{\bf{u}}(t')) dt' +
\int_{C} 
f^{\mu}_{\; ,i} (t',\mbox{\bf{x}}';\mbox{\bf{u}}(t')) dx'^{i} \; , 
\end{equation}
where $C$ is an arbitrary curve with constant $t'$, starting from $0$ 
and ending at $-A_{j}^{\; i}(t')x'^{j}$. 
The subintegral function in the second term  
of (\ref{el3}) is a total derivative, so this term does not depend on the 
curve $C$ and can be easily integrated. The time derivative in the 
first term is taken with $\mbox{\bf{u}}(t')$ kept fixed, so
$f^{\mu}_{\; ,0}$ in this term is not a total derivative.

Let us now apply the general formalism described in this section to a 
uniformly rotating ring. We assume that the ring is put in a rigid 
nonrotating circular gutter with the radius $R$, 
which provides that the radius of the rotating ring 
is the same as the radius of the same ring when it does not rotate, 
and is equal to $R$, 
as seen by an observer in $S$. This allows us not to worry about 
the complicated dynamical forces that tend to change the radius of the 
ring as seen by the observer in $S$, and pay all our attention to the 
kinematic effects resulting from the transformations
(\ref{er1})-(\ref{er2}). 

The ring can be considered as a series of independent short rods,
uniformly distributed along the gutter. (By a short rod we 
understand a rod with a length much shorter than $R$.) We assume 
that the gutter is placed at the $z=0$ plane. We put the 
space origin of $S$ at a fixed point on the gutter, such that 
the $y$-axis is tangential to the gutter and the $x$-axis is perpendicular 
to the gutter at $\mbox{\bf{x}}=0$. (In the rest of this section, as well as
in Sections 4 and 5, 
$\mbox{\bf{x}}\equiv
( x,y)$ and the $z$-coordinate is suppressed.)     
We study a single short rod initially placed at $\mbox{\bf{x}}=0$ and  
uniformly moving along the gutter in the counterclockwise direction.  
(This mimics a uniform motion of an electron in a 
synchrotron). The gutter causes a torque that provides that the rod is 
always directed tangentially to the gutter. Therefore,  
$\omega=u/R$, where $u=\sqrt{\mbox{\bf{u}}^2}$ is time independent.    
Now, $\gamma=1/\sqrt{1-\omega^2 R^2/c^2}$ is also time independent. 
Since a clock in $S'$ is at $\mbox{\bf{x}}'=0$, the clock rate between 
a clock in $S$ and a clock in $S'$ is given by $t=\gamma t'$, as seen 
by an observer in $S$. We assume that, initially, the axes $x'$, $y'$ are 
parallel to the axes $x$, $y$, respectively. Therefore the velocity
\begin{equation}
\mbox{\bf{u}}(t')=\omega R (-\sin \gamma\omega t', \cos \gamma\omega t') 
\end{equation}
is always in the $y'$-direction and the solution of (\ref{er4}) is 
\begin{equation}
A_{ij}(t')=\left( 
 \begin{array}{cc}
  \cos \gamma\omega t' & \sin \gamma\omega t' \\
  -\sin \gamma\omega t' & \cos \gamma\omega t'
 \end{array} \right) \; .
\end{equation}
The transformations (\ref{er1})-(\ref{er2}) become
\begin{equation}\label{er1'}
 \left( \begin{array}{c} x \\ 
                         y 
        \end{array} \right)=
 \left( \begin{array}{cc}
   \cos \gamma\omega t' & -\gamma\sin \gamma\omega t' \\
   \sin \gamma\omega t' & \gamma\cos \gamma\omega t'
  \end{array} \right)
 \left( \begin{array}{c} x' \\ 
                         y' 
        \end{array} \right)
 +R \left( \begin{array}{c} \cos \gamma\omega t' -1 \\ 
                            \sin \gamma\omega t' 
           \end{array} \right) \; ,
\end{equation}
\begin{equation}\label{er2'}
t=\gamma t' +\frac{\gamma}{c^2}\omega R y' \; .
\end{equation}
In particular, at $t'=0$ these transformations become
\begin{equation}\label{t=0}
 x=x' \; , \;\;\;\; y=\gamma y' \; , \;\;\;\; t=\frac{\gamma u}{c^2}y' \; , 
\end{equation}
which coincide with the ordinary Lorentz boost at $t'=0$ for the velocity in 
the $y$-direction.

\section{General coordinate transformations and the space 
line element in a non-time-orthogonal frame} 

A non-time-orthogonal frame is a frame in which $g'_{0i}$ is different 
from zero. It is generally accepted that the space line element in such a frame 
is given by (\ref{eq3}). However, if we assume that this formula can be 
applied to calculate the space distance as seen by a local observer,
then, as we have found in Section 1, Eq. (\ref{eq3})
leads to an imaginary length of a distant unaccelerated object as seen 
by a rotating observer. In order to resolve this puzzle, we examine the
assumptions under which formula (\ref{eq3}) is derived. 

In \cite{land}, formula (\ref{eq3}) is derived by assuming that the space 
distance between two points is measured by measuring the time $\Delta t'$ that 
light needs to travel from point $A$ to point $B$  and then back to   
point $A$. It is also assumed that the time is measured by a clock 
that does not change its position $x'^i$. 
The definition of the space distance $l'=c\,\Delta t'/2$ leads to 
(\ref{eq3}). 

In order to perform the described measurement in a 
rotating frame, the clock must be 
positioned at point $A$. However, according to our interpretation of 
(\ref{eq1}), this point can be faraway 
from the center of the rotation, so the required velocity 
of point $A$ can exceed $c$, as seen in $S$. Therefore, 
in general, such a measurement cannot be performed. 
%In other words, 
%(\ref{eq3}) does not always represent the space distance as seen by a local
%observer. 

In practice, we measure space distances between distant objects in a 
completely different way, namely, by measuring the angles under which we see the 
objects. (We assume that we know the radial distance of these objects from
us. The radial distance is not problematic in the theoretical sense, because 
$g'_{0r}=0$ in (\ref{eq2})).  
Our rotation does not influence this angle. Therefore, 
the apparent velocity of distant objects 
can exceed the velocity of light owing to our rotation, 
but a pure rotation (without velocity) 
will not lead to relativistic contraction, nor to elongation. The effect is 
that, in a rotating frame, it is more appropriate to calculate the space 
line element as 
\begin{equation}\label{dl} 
dl'^2=-g'_{ij}dx'^{i}dx'^{j} \; ,
\end{equation} 
despite the fact 
that $g'_{0i}$ is different from zero. This formula should be used 
to calculate the space distance between two arbitrary points 
which have the same $t'$ coordinate, no matter 
how far these points are from the observer at $x'^i =0$. 
Of course, if these points are end points of a body,    
then, in general, the distance calculated in this way 
will not be equal to the proper length of the body, but merely to 
the length seen by the observer.   
Formula (\ref{dl}) 
is also correct for 
frames that are both accelerated and rotating, defined by 
(\ref{er1})-(\ref{er2}).    
   
To clarify the meaning of formula (\ref{eq3}) completely, note that in 
\cite{mol} this formula is derived in a completely different way, 
without referring to any particular method of measurement. However, 
what is actually derived in \cite{mol} is the fact that the 
quantity (\ref{eq3}) does not change under coordinate transformations 
of the form
\begin{equation}\label{sameframe}
 t''=f^0 (t',x'^1,x'^2,x'^3) \; , \;\;\;\; x''^i=f^i (x'^1,x'^2,x'^3) \; .
\end{equation}
We refer to such transformations as {\em internal transformations}. 
Obviously, (\ref{eq1}) is not an internal transformation. 
Regular internal transformations form a subgroup of the group of all 
regular coordinate transformations.  
Note that the invariant quantity 
$ds^2=g_{\mu\nu}dx^{\mu}dx^{\nu}$ can always be written as 
\begin{equation}\label{logun}
ds^2=d\eta^2 -\gamma_{ij}dx^i dx^j \; ,
\end{equation}
where 
\begin{equation}\label{logun2}
d\eta^2 =\left[ \frac{g_{0\mu}dx^{\mu}}{\sqrt{g_{00}}} \right]^2 \; ,
\end{equation}
so $d\eta^2$ also does not change under internal 
transformations. The quantity $d\eta^2$ is nothing else but a time line element  
\cite{land}, defined by a measuring procedure similar to  
the measuring procedure used to define 
the space line element (\ref{eq3}). 

Let us illustrate the power of (\ref{eq3}), (\ref{sameframe}), and 
(\ref{logun2}) on the example that has already been discussed at some 
length in \cite{mol}. The Galilei transformation
$t''=t$, $x''=x-ut$ can also serve as a correct coordinate transformation
needed 
to describe the relativistic effects related to a frame moving with a constant
velocity $u$. The metric in these coordinates is given by 
\begin{equation}\label{mol1}
ds^2=c^2(1-u^2/c^2)dt''^2 -2udx''dt''-dx''^2 \; , 
\end{equation}
where it has been assumed that the metric of $S$ is given by 
$ds^2=c^2\, dt^2 -dx^2$. From (\ref{eq3}) and $dt=0$ one can 
obtain the relativistic contraction $dl=dx=dl''/\gamma$, 
where $\gamma =1/\sqrt{1-u^2/c^2}$. Similarly,   
from (\ref{logun2}) and $dx''=0$ one can obtain $dt=\gamma d\eta''$. 
The frame $S''$ is physically equivalent to the frame 
$S'$ which would be obtained from $S$ by the ordinary Lorentz
transformations, in the sense that $S''$ and $S'$ are connected by an
internal coordinate transformation 
\begin{equation}\label{mol2}
x'=\gamma x'' \; , \;\;\;\;\; t'=t''/\gamma -\gamma u x''/c^2 \; .
\end{equation}
Note that the non-time-orthogonal metric (\ref{mol1}), unlike (\ref{eq2}) and 
(\ref{metric}), can be transformed to a time-orthogonal metric by an 
{\em internal} transformation. Note also that the metric (\ref{mol1}), unlike
(\ref{eq2}) and (\ref{metric}), is not a metric of a Fermi frame. 

In \cite{mol}, internal transformations are interpreted as transformations 
that correspond to a redefinition of the coordinates of the same {\em
physical} observer. However,   
there is something unphysical about internal transformations; if $t'$ is a measure 
of the physical time for the observer in $S'$, then $t''$ is not, because 
it corresponds to a ``time" of the same observer which depends on the 
space point 
$x'^i$. Therefore, we introduce a more restrictive class of coordinate 
transformations, which could be better suited to interpret them  
as transformations that correspond to a redefinition of the coordinates of 
the same physical observer:
\begin{equation}\label{veryweak}
t''=f^0 (t') \; , \;\;\;\; x''^i=f^i (x'^1,x'^2,x'^3) \; . 
\end{equation}
We refer to such transformations as {\em restricted internal transformations}.
Regular restricted internal transformations form a subgroup of the group of all
regular internal transformations. 
The quantities $g'_{00}dt'^2$ and (\ref{dl}) do not change under 
restricted internal transformations.  

Now we have two definitions of the space line element,  
(\ref{eq3}) and (\ref{dl}), and related to this, two types 
of restricted coordinate transformations, internal and restricted internal. 
The space line element (\ref{eq3}) reduces to (\ref{dl}) if 
$g_{0i}=0$. However, as we have shown in this section,  
(\ref{dl}) is more appropriate in some cases, 
even if $g_{0i}\neq 0$. How to 
know in general what is the suitable definition of the 
space line element? 

We can immediately formulate one rule which is certainly 
suitable: {\em If the metric of a frame can be transformed to a 
time-orthogonal frame by an internal transformation, then the 
space line element should be calculated by (\ref{eq3})}. 

According to the results of this section, we can also formulate 
another rule: {\em If the metric of a frame in flat spacetime can be 
obtained from $g_{\mu\nu}=\eta_{\mu\nu}$ by a transformation of the form of
(\ref{er1})-(\ref{er2}) followed by an arbitrary restricted internal transformation, then 
the space line element should be calculated by (\ref{dl}).} 
Such coordinate transformations can be interpreted as the most general coordinate
transformations in flat spacetime that correspond to a physical observer
who has a positive mass. 

We still do not know a general rule. However, one can be satisfied 
to have a rule for Fermi frames only, or for Fermi frames modified 
by an arbitrary restricted internal transformation, because only such frames have a direct 
physical interpretation. One can be tempted to guess that 
for all such frames the space line element should be calculated by
(\ref{dl}), but such a conjecture requires further investigation.

For the sake of completeness, let us make a few remarks on general 
coordinate transformations in curved spacetime.  
The most general coordinate transformation 
that corresponds to a physical observer who has a positive mass is a 
transformation that leads 
to Fermi coordinates, followed by an arbitrary restricted internal transformation.  
Other coordinate frames may be useful for some physical calculations, 
for example, because it is easier to solve some covariant equations of motion 
in these coordinates. However, if one is interested in how 
the physical system  
looks like to a physical observer, one must transform the results to the 
coordinates specific for this observer. 

To summarize this section, we conclude that the correct definition 
of the space line element depends on how it is measured. Formula 
(\ref{eq3}) is not incorrect, but its applicability is limited and 
it should be used with great care. In our case of 
accelerated, rotating frames,  
it is more appropriate to calculate the space line element with 
$-g'_{ij}$ instead of with $\gamma'_{ij}$.

\section{Relativistic contraction}

In Section 2 we have found the coordinate transformation that describes 
the frame of a short rod uniformly moving along the circular gutter. 
Let as assume for a while that the length of 
the rod is infinitesimally small and 
that the rod is rigid (i.e., its proper length $dL'$ is equal to the 
proper length of the same rod when it does not accelerate).   
Let us determine the relativistic contraction of the rod, as seen by 
an observer in $S$. The observer in $S$ sees both ends of the rod
at the same instant, so $dt=0$.
From symmetry it is obvious that the relativistic contraction cannot depend on 
$t$, so, in order to simplify the calculations, we evaluate this at $t=0$. 
Since the rod is at $x'=y'=0$, (\ref{er2'}) implies that $t'=0$. Taking the
differential of (\ref{er1'}) and (\ref{er2'}) with respect to space and 
time coordinates, and then putting $x'=y'=t'=dt=0$, we find that the 
observer in $S$ sees the length
\begin{equation}\label{inf}
 dL=dy=\frac{dy'}{\gamma}=\frac{dL'}{\gamma} \; , 
\end{equation}
which is the expected relativistic contraction. 

Let us now turn our attention to the concept of the proper length of a 
body. Traditionally, it is defined as a length of the body as seen 
from the proper frame of the body. However, as we have seen, 
in general, there is no 
such thing as a proper frame of the body as a whole. Such a thing 
exists only for a nonrotating, inertially moving body in flat 
spacetime. The concept of a proper length 
of a large body does not have any fundamental meaning, 
simply because a ``large body" is not actually one object, but a 
set of many interacting particles. However, 
the proper length of an infinitesimally small part of a body is well 
defined. Therefore, we can define the proper length of a whole 
body as the sum of the proper lengths of its infinitesimal parts. 
Applying this to (\ref{inf}), we see
that the relativistic contraction of a short (but not infinitesimal) 
rigid rod uniformly moving along the circular gutter 
is given by $L=L_0/\gamma$, as seen by the
observer in $S$. Here 
$L_0$ is the proper length defined as above.  
 
Now, as in Section 2, assume that the rotating ring is a series of 
independent short rods, uniformly distributed along the gutter. 
Each rod is relativistically contracted, but the ring is not. 
This means that the 
distances between the neighboring ends of the neighboring rods 
are larger than those for a nonrotating ring, so  
the proper length of the ring is also larger than that of a nonrotating
ring. This is concluded 
also in \cite{gron2}. This situation mimics a more realistic 
ring made of elastic material, where atoms play the role of 
short rigid rods. Owing to the rotation the distances between 
neighboring atoms 
increase, so there are tensile stresses in the material. 
However, it is important to emphasize that the rotation is not 
essential for understanding of the origin of these tensile forces, 
because a similar effect also occurs in a linear relativistic 
motion \cite{dew}.   

The same relativistic contraction of  
short rods will be seen by a 
rotating observer in the center, because his frame is given by the 
Galilei transformation (\ref{eq1}) and the lengths are calculated by 
$g_{ij}$, as explained in Section 3. 
      
Let us now study how the nonrotating gutter looks like   
from the point of view of an observer on the rotating ring. 
Without 
losing on generality, we evaluate this at $t'=0$. We calculate the  
length of an infinitesimal part of the gutter lying near the 
observer, so $x=y=0$. Both ends are seen at the same instant, so 
$dt'=0$. Taking the
differential of (\ref{er1'}) with respect to space coordinates, 
and then putting $t'=0$, we find that the
observer in $S'$ sees the length
\begin{equation}\label{inf2}
 dL'=dy'=\frac{dy}{\gamma}=\frac{dL}{\gamma} \; , 
\end{equation}
which is the expected relativistic contraction. 

It is important to emphasize that (\ref{inf2}) is correct only in the 
infinitesimal form. The observer on the ring will not see other distant 
parts of the gutter contracted in the same way; for him, the gutter and 
the ring do not look azimuthally symmetric.  
In the following 
we study how other parts of the ring look like from the point of view 
of the observer on the ring. We introduce polar 
coordinates $r$, $\varphi$, defined by  
\begin{equation}
y=r \sin \varphi \; , \;\;\;\; R+x=r \cos \varphi \; ,
\end{equation}
which are new space coordinates for $S$, with the origin in the center 
of the circular gutter. The angle $\varphi$ is a good label of the position 
of any part of the ring even in $S'$. (To visualize this, one can draw  
angular marks on the gutter. The number of marks separating two points 
on the gutter or on the ring is a measure of the ``angular distance" in 
any frame.)   
Let $S''$ be the frame of another 
part of the ring. The position of that part of the ring is 
$x''=y''=0$. The relative position of the space origin of $S''$ with respect 
to that of $S'$ is given by the constant relative angle $\Delta\varphi_0$, 
as seen by an observer in $S$. In analogy 
with (\ref{er1'})-(\ref{er2'}), we find that $S''$ is determined by 
\begin{equation}\label{er1''}
 \left( \begin{array}{c} x \\
                         y
        \end{array} \right)=
 \left( \begin{array}{cc}
   \cos (\gamma\omega t''+\Delta\varphi_0) & -\gamma\sin (\gamma\omega
          t''+\Delta\varphi_0) \\
   \sin (\gamma\omega t''+\Delta\varphi_0) & \gamma\cos (\gamma\omega
          t''+\Delta\varphi_0)
  \end{array} \right)
 \left( \begin{array}{c} x'' \\
                         y'' 
        \end{array} \right)
 +R \left( \begin{array}{c} \cos (\gamma\omega t''+\Delta\varphi_0) -1 \\ 
                            \sin (\gamma\omega t''+\Delta\varphi_0) 
           \end{array} \right) \; ,
\end{equation}
\begin{equation}\label{er2''}
t=\gamma t'' +\frac{\gamma}{c^2}\omega R y'' \; .
\end{equation}
The observer in $S'$ will see the other part of the ring at the 
relative ``angular distance" $\Delta\varphi$, which, owing to the relativistic 
effects, differs from $\Delta\varphi_0$. Let the labels $A$, $B$ denote the 
coordinates of the part of the ring that lie at $S'$ and $S''$, 
respectively. Since the rotation is uniform, 
the relative ``angular distance"  
\begin{equation}\label{Dphi}
\Delta\varphi=\varphi_{B}(t''_{B})-\varphi_{A}(t''_{A}) 
 = \Delta\varphi_0 +\gamma\omega t''_{B}-\gamma\omega t'_{A} \; ,
\end{equation}
cannot depend on $t'$, so without losing on generality, we evaluate this at
$t'=0$. Since the observer sees both parts of the ring at the same 
instant, we have $t'_{A}=t'_{B}=0$. Since $x''_{B}=y''_{B}=0$, from 
(\ref{er1''}) we find
\begin{equation}\label{y2}
y_{B}=R\sin (\gamma\omega t''_{B}+\Delta\varphi_0) \; ,
\end{equation}
and from (\ref{er2''}) 
\begin{equation}\label{t2}
t_{B}=\gamma t''_{B} \; .   
\end{equation}
From $t'_{B}=0$ and (\ref{t=0}) 
it follows $t_{B}=\omega R y_{B}/c^2$, which, because
of (\ref{t2}), can be written as $\gamma t''_{B}=\omega R y_{B}/c^2$. 
This, together with (\ref{y2}), leads to the equation that determines
$t''_{B}$:
\begin{equation}\label{eqgron1}
\gamma\omega t''_{B}=\beta^2\sin (\gamma\omega
t''_{B}+\Delta\varphi_0) \; , 
\end{equation}
where $\beta^2 \equiv \omega^2 R^2 /c^2$. 
From $t'_{A}=0$ and (\ref{Dphi}) we see that $\Delta\varphi=\gamma\omega
t''_{B}+\Delta\varphi_0$, so (\ref{eqgron1}) can be written as 
\begin{equation}\label{eqgron1'}
\Delta\varphi -\Delta\varphi_0 =\beta^2\sin \Delta\varphi \; .
\end{equation}

Equation (\ref{eqgron1'}) determines the relative ``angular 
distance" $\Delta\varphi$ 
between two points on the ring as seen by the observer at one of the 
points, if the relative angle between these two points, as seen by the
observer in $S$, is $\Delta\varphi_0$. In other words, (\ref{eqgron1'}) determines 
how the ring looks like to the observer on the ring. For an inertial 
observer whose instantaneous 
position and velocity are equal to that of the observer on the ring, 
the same equation (\ref{eqgron1'}) is found in \cite{gron1}, where 
the solution is graphically depicted. This means, contrary to the 
conclusion of \cite{gron1}, that the inertial and the noninertial 
observers see the ring in the same way.   

If the two points on the ring are very close to each other, then 
$\Delta\varphi_0$ and $\Delta\varphi$ are very small. By expanding 
equation (\ref{eqgron1'}) for small angles we find the approximative 
solution $\Delta\varphi=\gamma^2 \Delta\varphi_0$. The factor $\gamma^2$ is 
easy to understand; one factor of $\gamma$ appears because the 
part of the gutter close to the observer on the ring looks shorter 
for that observer than it really is, and the other factor of $\gamma$ 
appears because the part of the ring close to the observer on the ring  
is longer than that of the same ring when it does not rotate. 

\section{The rate of clocks}

Assume that there are two clocks at different positions on the 
ring. Assume also that they show the same time, as seen by an 
observer in $S$. Then, as shown in Section 2, both clocks 
show the time $t'=t/\gamma$, as seen from $S$. 

These two clocks do not show the same time as seen by an observer 
on the ring. If the position of the observer coincides with the 
position of one of the clocks, then the time-shift of the other 
clock is given by (\ref{eqgron1}). 

Let us calculate the time-shift of the clock at the fixed position 
$(x,y)$, as seen by the observer in $S'$. From (\ref{er1'}) 
we express $y'$ as a function of $x$, $y$, and $t'$, and put this 
in (\ref{er2'}). The result is 
\begin{equation}\label{txy}
t=\gamma t' + 
 \frac{\omega R}{c^2}[y \cos \gamma \omega t' 
 -(x+R) \sin \gamma \omega t'] \; .
\end{equation}
For comparison, if (\ref{er1'}) and (\ref{er2'}) are replaced by 
the ordinary Lorentz boosts for a constant velocity in the $y$-direction, 
then (\ref{txy}) should be replaced by 
\begin{equation}\label{txyo}
t=\frac{t'}{\gamma}+\frac{u}{c^2}y \; . 
\end{equation}

To understand the physical meaning of (\ref{txy}), 
we explore some special cases. 
If $\gamma \omega t'=2k\pi$, then $t=\gamma t'+\omega R y/c^2$. In this  
case, the rate of clocks $\Delta t/\Delta t'=\gamma$ is the same as 
that for the observer 
in $S$. This can also be understood as a time-averaged rate, 
because the oscillatory functions in (\ref{txy}) vanish when 
they are averaged over time. Therefore, the observer in $S'$ agrees 
with the observer in $S$ that the clock in $S'$ is slower, but only in 
a time-averaged sense. 
At some instants the observer in $S'$ sees that 
the clock in $S$ is slower than his clock.   
For example, by putting $x=0$ and 
expanding (\ref{txy}) for small $t'$, we recover formula (\ref{txyo}), 
with $u=\omega R$. If the clock in $S$ is in the center, which corresponds 
to $x=-R$, $y=0$, then (\ref{txy}) gives $t=\gamma t'$, so in this case 
there is no oscillatory behavior.    

\section{Velocity of light}

Let us also make some comments on the velocity of light. The Sagnac effect
is usually interpreted as a dependence of the velocity of light
on the direction of light propagation in a rotating frame
(see, for example, \cite{post}, \cite{tart} and references therein).
However, such an interpretation is based on the interpretation
of the frame $S'$ defined by (\ref{eq1}) as a proper frame of all
observers on a rotating platform. Now we know that each observer    
belongs to a different local Fermi frame, and from (\ref{metric}) we see that
in the {\em vicinity} of any observer the metric is equal to the
Minkowski metric $\eta_{\mu\nu}$. This implies that {\em
for any local observer the velocity of light is isotropic and is
equal to $c$, providing that it is measured by propagating a light
beam in a {\bf small} neighborhood of the observer, 
using Einstein synchronized clocks}. This is
also true for an observer in curved spacetime, because his
proper frame is given by the appropriate Fermi coordinates, which
also have a property that $g_{\mu\nu}=\eta_{\mu\nu}$ at the position
of the observer. The phrases
``local" and ``small" denote spatial dimensions inside which
the metric tensor does not change significantly.

Of course, the velocity
of light does not have to be equal to $c$ for an observer which is not
at the same position as the light. However, this is not only a property
of non-time-orthogonal frames. For example, if the acceleration
of an uniformly accelerated observer and the propagation of light
are both in the $x'$-direction, then from (\ref{metric}) one can
find that the accelerated observer sees the velocity of light as  
$|dx'/dt'|=c\sqrt{1+a'x'/c^2}$, being equal to $c$ only at     
$x'=0$. A similar effect occurs for a radial 
motion of light in the vicinity of the Schwarzschild radius of a black hole,
as seen by a static observer faraway from the Schwarzschild radius.

Concerning the Sagnac effect, we do not claim that the standard prediction  
for the phase shift is incorrect. It can also be derived by performing
calculations in the nonrotating frame $S$ \cite{post}, and such a 
derivation, based on the well-understood Minkowski 
spacetime, is perfectly correct. We have nothing new to say about 
the phase shift, which appears when clockwise and counterclockwise
propagated  
light beams finally meet. However, as seen by an observer on the rim of a 
rotating disc, the velocity of the light beam will be a complicated function 
of time $t'$, or equivalently, of the position $(x',y')$ of the beam. 
The trajectory of 
the light beam expressed in $S$-coordinates takes a simple form 
\begin{equation}\label{sag}
y=R \sin \omega_L t \; , \;\;\;\;\; 
x=R(-1+\cos \omega_L t) \; ,
\end{equation}
where $\omega_L =\pm c/R$. The plus and minus signs refer to the 
counterclockwise and clockwise propagated beams, respectively. 
Using (\ref{er1'}), (\ref{er2'}), and (\ref{sag}), one can eliminate 
$x,y,t$ and express $x',y'$ as functions of $t'$. The speed of 
light as seen by the observer in $S'$ is 
\begin{equation}\label{sag2}
v'_L =\sqrt{\left( \frac{dx'}{dt'}\right)^2 +
\left( \frac{dy'}{dt'}\right)^2 } \; .
\end{equation} 
Expanding (\ref{er1'}) and (\ref{sag}) for small $t'$ and $t$, respectively, 
one can easily find $y'=\pm ct'+{\cal O}(t'^2)$, $x'={\cal O}(t'^2)$, 
which means that the observer sees the velocity
of light equal to $c$ when the light is at the same position 
as the observer, just as expected.

\section{Discussion} 
 
From the experience acquired by careful calculations in the preceding  
sections, we can generalize some of the results without much effort, 
using qualitative and intuitive arguments. 

If an observation in $S$ is performed at the instant $t$, then the 
solution of (\ref{er4}) can always be chosen such that at $t$ 
the axes $x'^i$ are parallel to the corresponding axes $x^i$. Therefore,  
for a small range of values of $t'$, the transformations 
(\ref{er1})-(\ref{er2}) can be approximated by the ordinary Lorentz boosts (see
(\ref{t=0})). 
From this fact we conclude that   
if a moving rigid body is short enough, then its relativistic contraction   
in the direction of the instantaneous velocity, as seen from $S$, 
is simply given by $L(t)=L'/\gamma(t)$, i.e., it depends 
only on the instantaneous velocity, not on its acceleration and rotation. 
(``Short enough" means that $L'\ll c^2/a'_{\|}$, where $a'_{\|}$ is the 
component of the proper acceleration parallel to the direction 
of the velocity \cite{nikol}). 

By a similar argument we may conclude that 
an arbitrarily accelerated and rotating 
observer sees equal lengths of other differently 
moving objects as an inertial observer whose
instantaneous position and velocity are equal to that of the 
arbitrarily accelerated and rotating observer.  

So far we have studied a rotating ring. A rotating disc is a more 
complicated object, with some additional dynamical effects  
related to elastic and inertial forces. However, a disc can be modeled 
as a series of concentric rings, each of them being constrained to have a 
fixed radius. In this case, the analysis of a rotating disc becomes 
essentially the same as that of a rotating ring.  

Let us also give some additional arguments why our resolution of the 
Ehrenfest paradox is correct. Our method, based
on coordinate transformations (\ref{er1})-(\ref{er2}), is really 
a generalization of the well-known derivation of the Lorentz contraction 
for constant velocities. In our 
approach the origin of the relativistic contraction lies in 
the non-Galilean transformation, not in the nontrivial metric, whereas 
in the standard approach the transformation is Galilean and  
the contraction is due to the nontrivial metric (\ref{eq2}). 
Note finally that our approach allows a generalization to a more 
complicated motion, whereas the standard approach does not. 

Finally, let us make some comments on the observability of the relativistic 
contraction. In principle, it could be observed by photographing a rod 
with a very short exposition, such that both ends are observed 
at the same instant. Since the velocity of the incoming information 
(velocity of light) is finite, both ends of the rod should be 
positioned at the same distance from the observer. Therefore, the ideal 
setup for such a measurement is a rod in a uniform circular 
motion and a camera in the center, providing that we can achieve a 
short enough exposition. It is assumed that in this experiment the 
only object that moves circularly is a rod (with two ends); there is  
neither a rotating disc, nor a rotating ring.   

An indirect, but easier-to-perform experimental 
verification of the relativistic contraction could perhaps be obtained 
by measuring the velocity of a rotating ring in a rigid circular 
gutter, needed to achieve the break of the ring, and comparing it with 
the elongation needed to achieve the break of the ring 
caused by ordinary stretching. 

Of course, in both types of experiments 
the problem is to achieve a relativistic velocity
of macroscopic objects, so these can be 
considered merely as {\it gedanken} experiments.      

\section{Conclusion}

In this paper a new resolution of the Ehrenfest paradox has been 
provided by taking into consideration the fact that although there is no
relative motion
among different points on a rotating disc, each point belongs to a 
different noninertial local Fermi frame. 
If a rotating ring (or a disc) is constrained to have a
 fixed radius from the point of view of an inertial observer, it has been 
 found that there are tensile stresses in the disc, in agreement with the
 prediction of the standard approach. However, contrary to the
 prediction of the standard approach, it has been found that an
 observer on the rim of the disc will see equal lengths of other differently 
moving objects  
 as an inertial observer whose instantaneous
 position and velocity are equal to that of the observer on the rim, 
providing that the observations of different events are simultaneous. 
This also generalizes to observers arbitrarily moving in  
flat spacetime.   

The paper deals mainly with flat spacetime, with particular 
attention paid to circular motion. However, it gives several 
results which are of very general relevance, not only for arbitrary 
motion in flat spacetime, but also for general relativity and  
curved spacetime. 

First, it has been demonstrated that 
the generally accepted formula (\ref{eq3}) is not always correct. 
The correct definition of the space line element depends on how it 
is measured, so (\ref{eq3}) should be used with great care. 
In some cases, the ``naive" formula (\ref{dl}) is more appropriate. 
One such case is a metric of a frame in flat spacetime that can be
obtained from $g_{\mu\nu}=\eta_{\mu\nu}$ by a transformation of the form of
(\ref{er1})-(\ref{er2}), followed by an arbitrary restricted internal transformation. 
Further investigation is needed in order to generalize this result.

Second, the paper demonstrates the importance of the use of  
Fermi coordinates. One of the consequences of their use is the result 
that for any local observer the velocity of light is isotropic and is
equal to $c$, providing that it is measured by propagating a light
beam in a small neighborhood of the observer. 
This fact should be 
used for a correct treatment of the Sagnac effect 
if one wants to explore the general relativistic corrections. 
Fermi coordinates should also be used in order to understand the 
physical effects related to a rotating black hole, to give 
a correct treatment of the Hawking radiation, as well as 
for any other 
physical effect, whenever intended to describe the world 
how it looks like to a particular observer.    

\section*{Acknowledgment}

The author is grateful to Damir Stoi\'{c} for motivating 
discussions.
This work was supported by the Ministry of Science and Technology of the
Republic of Croatia under Contract No. 00980102.

\end{document}